\documentclass   [a4paper,12pt]{article}
\usepackage {graphicx,amsmath}
\begin{document}
\begin{center}
 {\Large\bf  Optical theorem and elastic nucleon scattering } \\[2mm]
            Milo\v{s} V. Lokaj\'{\i}\v{c}ek,  Vojt\v{e}ch Kundr\'{a}t    
\footnote{e-mail: lokaj@fzu.cz, kundrat@fzu.cz} \\
Institute of Physics of the AS CR, v. v. i., 182 21 Prague 8, Czech Republic  \\
\end{center}

{\bf Abstract } \\
In the theoretical analysis of high-energy elastic nucleon scattering one starts commonly from the description based on the validity of optical theorem, which allows to derive the value of total cross section directly from the experimentally measured t-dependence of elastic differential cross section. It may be shown, however, that this theorem has been derived on the basis of one assumption that might be regarded perhaps as acceptable for long-range (e.g., Coulomb) forces but must be denoted as quite unacceptable for finite-range hadron forces. Consequently, the conclusions leading to the increase of total cross section with energy at higher collision energies must be newly analyzed. The necessity of new analysis concerns also the derivation of elastic scattering t-dependence at very low transverse momenta from measured data.
 

\section{Introduction}
\label{sec1}
It is commonly argued that the total cross section of two colliding nucleons increases significantly with energy at higher energy values. However, the corresponding phenomenon has been derived from measured elastic differential cross section on the basis of the validity of optical theorem. And the given conclusion must be impeached since the optical theorem has been derived from the assumption that might be taken perhaps as admissible for long-range forces but is quite unacceptable for finite-range hadron interactions.
Consequently, it is not possible to derive the value of total cross section directly from the mere measured elastic differential cross section.

 In the following we shall analyze the common approach of deriving optical theorem and specify the corresponding assumption. The given approach should be denoted as unacceptable for finite-range hadron forces since in identifying the original incoming state (remaining unchanged in collision process) with one special state of scattered spectrum the unjustified assumption concerning scattering function has been applied to.

\section {Elastic collisions and optical theorem }
\label{sec2}

In quantum-mechanical approach the evolution of a two-particle system is described with the help of function $\psi(X,\tau)$ (solution of Schr\"{o}dinger equation) and elastic collision process has been interpreted standardly as the transition from the state at time $\tau=-\infty$ to the state at time $\tau=+\infty$, which have been represented by different vectors in the Hilbert space spanned on the vector basis formed by the set of Hamiltonian eigenfunctions. The initial state is represented commonly by plane wave $\psi_{in}=e^{ikz}$ and the final state is then expressed as the superposition of the set of individual scattered states.

As to the derivation of the optical theorem we shall follow the approach presented, e.g., in Ref. \cite{baro} that has started from Fraunhofer diffraction where the hole has been characterized by profile function $\Gamma(\mathbf{b})$. Then Babinet's principle has been applied, stating that a hole and an obstacle of identical form and dimension produce the same diffraction pattern. One can write  
\vspace{-3mm}
   \[  U(x,y.z)=-\frac{ik}{2\pi}\,U_0\,\frac{e^{ikr}}{r} 
    \int d^2{\mathbf b}\,S({\mathbf b})\,e^{-i{\mathbf q}.{\mathbf b}} \]
where $S({\mathbf b})$ characterizes obstacle; and similarly for hole if $S({\mathbf b})$ is substituted by $\Gamma(\mathbf{b})$. And finally, the Huygens-Fresnel principle has been applied to: the waves diffracted by the hole and by the obstacle of the same dimensions combine to reconstruct the incident plane wave; i.e. $S(\mathbf{b})+\Gamma(\mathbf{b})=1$. The whole approach being based on the assumption of small diffraction angles ($\sin\theta\cong\theta$).

One can then write
\vspace{-3mm}
\begin{equation}
  U(x,y,z) = U_{unsc} + U_{scatt}
  = U_0\biggl(\psi_{unsc}+f(\mathbf{q})\frac{e^{ikr}}{r}\biggr)  \label{ux}
\end{equation}
where $\mathbf{q}=\mathbf{k}'-\mathbf{k}$ is transversal momentum and $|\mathbf{k}'|=|\mathbf{k}|=k$; the modulus squared of $U(x,y,z)$ represents corresponding probability and $U_0$ - original intensity; $\psi_{unsc}$ representing here the non-interacting part of original state. The outgoing scattered states are  characterized by vectors  $\mathbf{q} \;(|\mathbf{q}|\in(0,q_{max})$ where $q_{max}$ is maximal possible value in  given collision case; depending on the energy of colliding objects.

 It is possible to write 
\vspace{-3mm}
\begin{equation}
     f(\mathbf{q}) = \frac{ik}{2\pi}\int\!\! d^2\mathbf{b}\;\Gamma(\mathbf{b})\;
                                 e^{-i\,(\mathbf{q}.\mathbf{b})}   \label{fq}
\end{equation}
where $\Gamma(\mathbf{b})$ is corresponding profile function. The differential cross section equals then 
       \[ \frac{d\sigma}{d\Omega}=|f(\mathbf{q})|^2 \]
and integrated elastic cross section equals (in approximation for small scattered angles $\theta\cong\sin\theta$)        
\vspace{-2mm}
\begin{equation}
    \sigma_{el}=\frac{1}{k^2}\int|f(\mathbf{q})|^2d^2\mathbf{q}
    \cong  \int\!\! d^2\mathbf{b}\,|\Gamma(\mathbf{b}|^2 =
                   \int\!\! d^2\mathbf{b}\,|1-S(\mathbf{b})|^2 \,  .     
\end{equation}
If $\psi_{unsc}$ is identified with the state characterized by $q=0$ and the total interaction is normalized to unity the absorption cross section is given by
\begin{equation}
   \sigma_{abs}= \int\!\! d^2\mathbf{b}[1-S^2(\mathbf{b})] =
   \int\!\! d^2\mathbf{b}[2Re\Gamma(\mathbf{b})-|\Gamma(\mathbf{b})|^2]  \,. 
\end{equation}
And
\begin{equation}
     \sigma_{tot}=\sigma_{el}+\sigma_{abs}=
            2\int\!\! d^2\mathbf{b}\,Re\Gamma(\mathbf{b})  \,.   \label{sitot}
\end{equation}
When Eq. (\ref{fq}) is taken into account and one other assumption (or rather two assumptions - see below) is added it may be deduced from Eq. (\ref{sitot})
\begin{equation} 
     \sigma_{tot} = \frac{4\pi}{k}\mathrm{Im}f(\mathbf{q}=0)   \label{sgt}
\end{equation} 
In deriving Eq. (\ref{sgt}) the non-interacting state (see Eq. (\ref{ux})) has been identified with the outgoing state characterized by ${\mathbf q}=0$. And the condition of unitarity has been applied to reduced outgoning state. 

However, let us see what has been actually done: Eq. (\ref{ux}) has been substituted in principle by 
\begin{equation}
  U(x,y,z) = U_0\;F(\mathbf{q})\frac{e^{ikr}}{r}   \label{ux2}
\end{equation}
where $F(\mathbf{q})$ has differed from $f(\mathbf{q})$ only in one point: $q=0$; holding  
    $$   F(0)= f(0) + |\psi_{unsc}|. $$ 
One singular point has been added to the function $f(\mathbf{q})$ and the condition of unitarity has been applied to the function $F(\mathbf{q})$; the approach being in principle acceptable until now.
However, what has followed might be probably acceptable for long (infinite)-range forces, but it must be denoted surely as unacceptable fort finite-range (contact) forces. 

Two assumptions concerning the function $F(\mathbf{q})$ have been added in principle:  

 - function $F(\mathbf{q})$ has been assumed to be continuous  even if the changed value in point $q=0$ has been included;  

 - function $F(\mathbf{q})$ has been taken as a rule as decreasing (or at least non-increasing) with rising $q$ in the neighborhood of $q=0$.

From the first assumption it follows $|\psi_{unsc}|=0$, which is surely in contradiction to experimental facts in the case of nucleon collisions when fully unscattered nucleons are contained in the range of measured beam. And there is not any reason for the assumption of non-rising function $F(\mathbf{q})$ (or $f(\mathbf{q})$), either. Both the assumptions are, therefore,  unacceptable (and non-reasoned) in the case of hadronic collisions, even if they might be  probably admitted for infinite-range Coulomb force (at least to some extent). There is surely a qualitative difference between these two kinds of forces as to the validity of optical theorem.

There are, therefore, some questions that are open now. It is the actual shape of function $f(\mathbf{q})$; especially, the value of $f(0)$ is very important. It is also the problem of estimation of colliding beam luminosity from elastic scattering that remains open now.  

\section{Collisions at very small angles}
\label{sec3}

At very small angles the differential hadronic cross section is fully hidden under the Coulomb effect. And if the optical theorem cannot be made use of the corresponding $t$-dependence of nucleon anplitude represents fully open question.

In the standard approach the increase of $Imf(0)$ with energy is, however, evidently necessary without having any considerable physical impact. It may be a direct consequence of the mere narrowing of diffraction peak. The usual complete amplitude corresponding approximately to energy of 53 GeV is represented by full line in Fig. 1. It is given as the sum of Coulomb and hadronic amplitudes; the influence of relative phase and of the real part of hadronic amplitude being neglected now. The curve connected with the full line by vertical line represents then the hadronic amplitude when both the parts (Coulomb and hadronic) are mutually equal at $t=-0.02\,GeV^2$.

The other pair of curves in Fig. 1 represents then the behavior at higher energy when the hadronic peak should be expected to be much narrower. It is evident that the hadronic value at $t=0$ must rise even if integral elastic cross section (and, eventually, total cross section) might be significantly lower than in the first case.

Consequently, there is also the problem with  estimation of beam luminosity from elastic scattering data, if the narrowing of hadronic part is not taken correspondingly into account. The luminosity estimation may be easily systematically undervalued in comparison to any preceding case of lower energy and, consequently, all physical characteristics measured at higher energy may be overvalued. However, the problem of luminosity undervaluation will be seen better from the next figure.

\begin{figure}[t!]
\begin{center}
\includegraphics*[scale=.40, angle=-90]{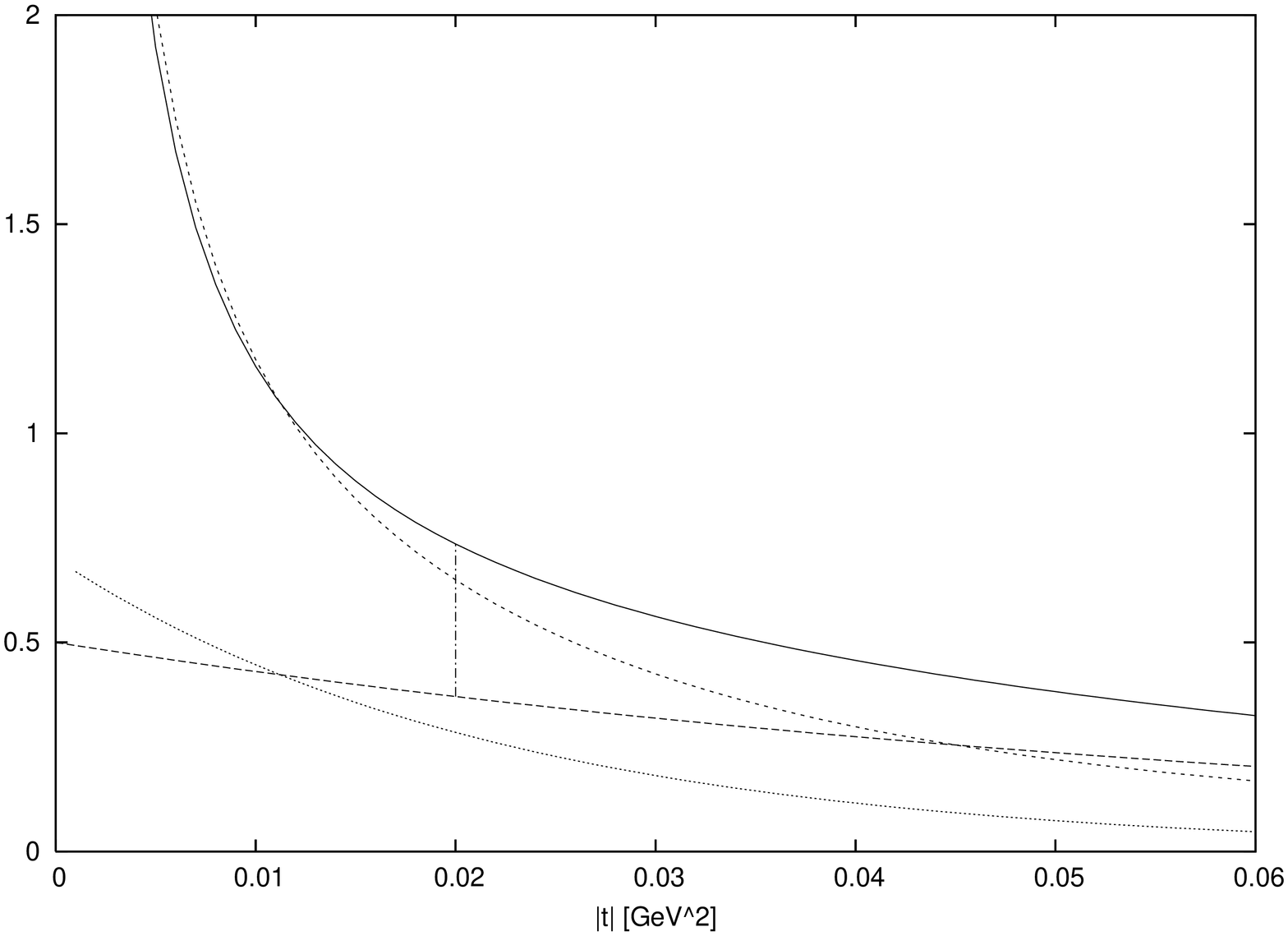}
\caption { \it {Comparison of hadronic amplitudes at different energy values  } }
\vspace{4mm}
\includegraphics*[scale=.40, angle=-90]{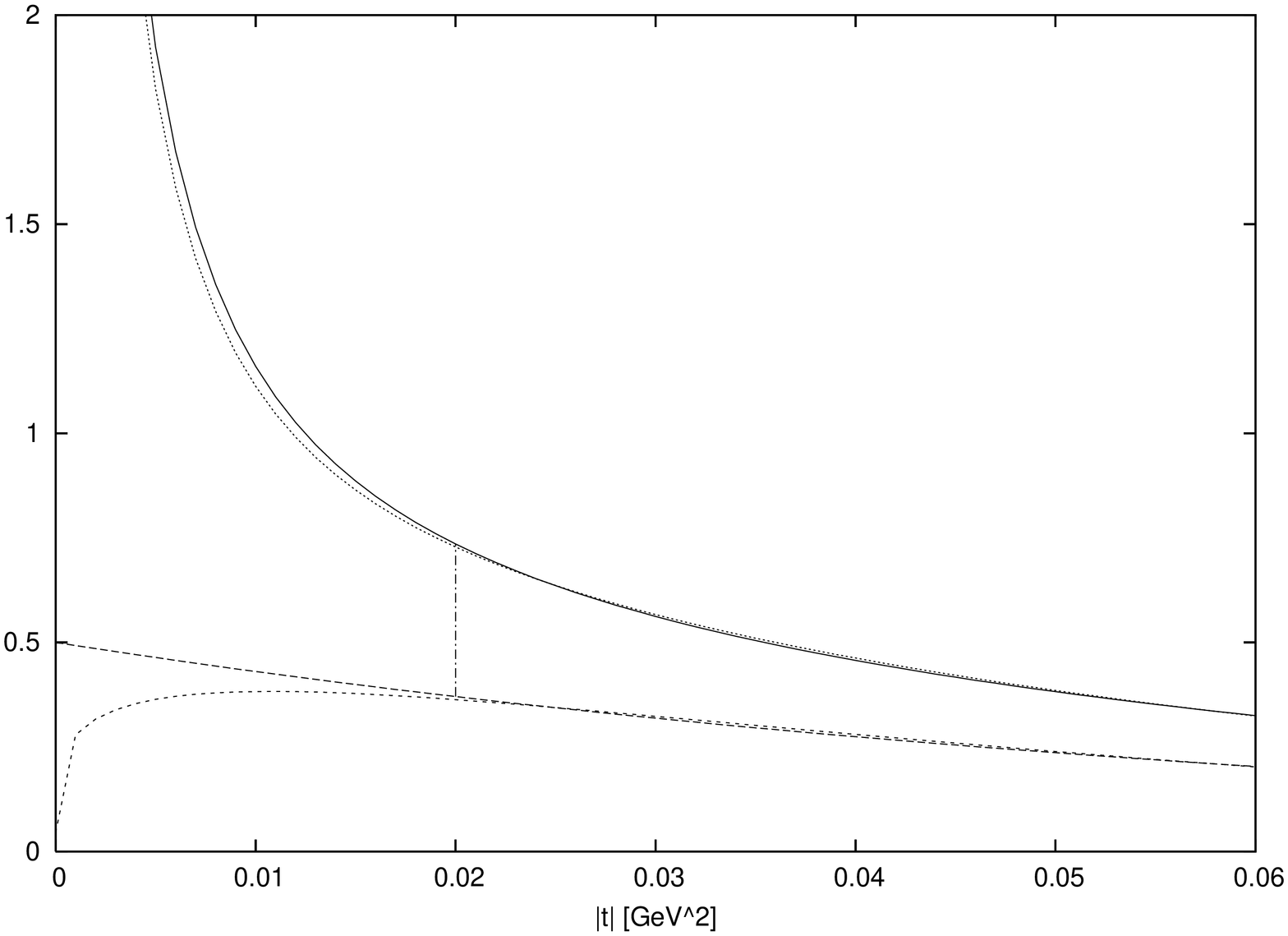}
\caption { \it {Different hadronic amplitudes at 53 GeV } }
 \end{center}
 \end{figure}

If the optical theorem is not to be made used of, the purely phenomenological approach is quite insufficient in establishing the $t$-dependence of hadronic amplitude at very small angles; see Fig. 2. The first pair of lines reproduces again the standard results at 53 GeV. And it is evident from the second pair that practically the same complete line may be obtained, even if the t-dependence of hadronic amplitude at very small collision angles is quite different. And new ways should be looked for to solve preceding problems.  

\section{Conclusion}
\label{sec4}

It follows from the preceding analysis that the contemporary phenomenological theories of elastic collisions do not allow to establish the total cross sections without some additional experiments or without more detailed analysis based on a model in which elastic and inelastic processes would be mutually correlated on realistic physical grounds. 
There is also the question how it is with the interpretation of nucleon elastic data for very small values of $q$. Are they really decreasing from the value at $q=0$ or may they exhibit different behavior being fully hidden under the effect of Coulomb force? The problem is related very closely also to establishing luminosity value from elastic scattering data.

Important problem must be seen also in the fact that the $t$-dependence of elastic differential cross section is the only available experimental set of data from which the mere modulus of complex amplitude function may be established while the phase remains practically undetermined. Quite different impact-parameter characteristics may be then derived according to its choice; see, e.g., \cite{vkun,y1981}. It means that ontologically very different behaviors may be derived on the basis of standard phenomenological models.

{\footnotesize

\end{document}